\documentclass{jps-cp}
\usepackage{txfonts} 

\title{Uncorrelated Far Active Galactic Nuclei Flaring With Their Delayed Ultra High Energy Cosmic Rays Events}

\author{Daniele \textsc{Fargion}$^{1,2,3}$, Pietro \textsc{Oliva}$^{3,4,5}$ and Pier Giorgio \textsc{De Sanctis Lucentini}$^{6}$}

\inst{
$^{1}$Physics Department, Rome University 1, P.le A. Moro 2, 00185, Rome, Italy \\
$^{2}$INFN Rome1, P.le A. Moro 2, 00185, Rome, Italy \\
$^{3}$Mediterranean Institute of Fundamental Physics, Via Appia Nuova 31, 00040 Marino (Rome), Italy \\
$^{4}$Engineering Dept., Niccol\`o Cusano University, Via Don Carlo Gnocchi 3, 00166 Rome, Italy \\
$^{5}$Department of Sciences, University Roma Tre, Via Vasca Navale 84, 00146 Rome, Italy \\
$^{6}$Physics Department, Gubkin Russian State University
of Oil and Gas (National Research University), 65 Leninsky Prospekt, Moscow, 119991, Russian Federation
}

\email{daniele.fargion@uniroma1.it}

\recdate{}

\abst{The most distant Active Galactic Nuclei (AGN), within the allowed Greisen-Zatsepin-Kuzmin (GZK) cut-off radius ($\lesssim100$~Mpc), have been recently
candidate by many authors as the best location for the observed Ultra High Energy Cosmic Rays (UHECR) origination.
Indeed, the apparent homogeneity and isotropy of recent UHECR signals seems to require a far cosmic isotropic and homogeneous scenario, involving a proton UHECR courier: our galaxy or nearest local group or super galactic plane (ruled by the Virgo cluster) are too near and apparently too anisotropic to be in agreement with the (Pierre Auger Observatory (PAO) and Telescope Array (TA) almost-homogeneous  data sample.
However, the few and mild UHECR observed clustering, the so called North and South Hot Spots, are smeared in wide ($\pm18^\circ$) solid angles. Their consequent random walk flight from most far GZK UHECR sources, nearly at 100~Mpc, must be delayed -- with respect to a straight AGN photon gamma flaring arrival trajectory -- at least by a million years.
During this time, the AGN jet blazing signal, its probable axis deflection (such as the helical jet in Mrk~501),
its miss alignment or even its almost certain exhaust activity, may lead to a complete misleading correlation
between present UHECR events and a much earlier active AGN ejection. UHECR maps may be anyway related to  galactic or nearest (Cen~A, M82) AGN extragalactic UHECR sources shining in twin Hot Spot. Therefore we defend
our (quite  different) scenario where UHECR are mostly made by lightest UHECR nuclei originated by nearby AGN sources, or few galactic sources, whose delayed signals are reaching us within few thousand years in the observed smeared sky areas.}

\kword{cosmic rays, UHECRs,  AGN, intergalactic magnetic fields}

\begin{document}
\maketitle

\section{Introduction}
The Cosmic Ray (CR) puzzle is, since Hess discover a century ago, a  scientific unsolved key problem in particle and astrophysics science.
The mysterious absence of  magnetic monopoles (unobserved in laboratory) reflects into the survival and the presence of large size magnetic fields along stellar, galactic and cosmic spaces; these magnetic fields (measured by polarization due to faraday rotation of radio sources) may bend by Lorentz forces the charged CR, protons and nuclei, making them totally deflected and smeared while raining on the Earth, with little or no arrival direction-source correlation.
Moreover, the presence of a cosmic microwave black body photon bath makes the most energetic CR events, the UHECR at tens of EeV energy edges, bounded within a very narrow cosmic radius: the so called GZK cut off \cite{GZK(1966)}.
Their hunt would be restricted in a limited volume and it might be an easy goal to achieve. Indeed these energetic UHECR, if nucleons, would be so rigid, while in flight, to get hard to be bent by galactic or cosmic Lorentz forces, and thus they would be able to keep at least a coarse memory (which translates into anisotropy) of their origination.
Therefore, the hope was, since the discover of the UHECR, to track their maps in the sky possibly correlating them with the best astrophysical $\gamma$ sources as Super-Nov\ae{} (SN), Gamma Ray Bursts (GRB), BL~Lac, AGN or just the nearest micro galactic or macro extragalactic jet in our neighbour Universe.
In general the hope was to connect highest energy gamma spots and highest energy $\nu$ maps with these UHECR events. Also the mass distribution (the Local Group, the Super Galactic Plane -- SGP) was expected to tag the AGN and the UHECR sources. None of these obvious results have been achieved. One may wonder why we should care and despair: the reason is that the inquire for the UHECR sources, or their smoking guns, would answer to CR origination places, offering a deep  understanding and a view for the most energetic astrophysical events.

Since 25 years and up to now, after the early Fly's Eye event \cite{Elbert(1995)} (as well as the maximal, more and more puzzling, unbeaten  $3\cdot10^{20}$~eV energetic event), after Akeno Giant Air Shower Array (AGASA) decade of records\cite{1999ApJ...510L..91M, 2000AGASA}, the HiRes observation of a UHECR cut off \cite{2007HiRes}, and the more recent hundreds of PAO and TA  records \cite{Kampert16, PAO16, 2016JCAP}, some (or let say most) of the UHECR are still mainly spread.
In summary on  years 1991-2000, up to HiRes results, we were all believing that  UHECR could overcome somehow the GZK cut off: this is because some of the most energetic \cite{Elbert(1995)} UHECR were uncorrelated with nearby (GZK distance) sources and because AGASA didn't show the GZK cut off.
To face this over-GZK possibility few of us \cite{Fargion1999} suggested the presence of relic dark neutrino halos with a rest eVs masses, able to convert far AGN UHE ZeV neutrino energy by $\nu_{\mathrm{ZeV}}+\bar{\nu}_{\mathrm{relic}}$ scattering  within a few Mpc ${\nu}_{\mathrm{relic}}$ halo, with energy at center of mass at Z boson resonance value; the Z ultra-relativistic decay may eject later on $p$ ,$\bar{p}$, $n$, $\bar{n}$ secondaries that could finally shine and shower on Earth's atmosphere as UHECR from far (above GZK cut off) Universe edges.
Highest energy ZeV neutrino have the virtue of flying from any far Universe edges almost undisturbed by the (GZK) cut off (due to 2.75 K BBR microwave photons) but they could hit and convert their energy by annihilation while scattering onto relic cosmic antineutrino target (spread in huge warm dark halos) though in a rare tuned resonance energy mass value of $E_{\nu}\simeq4\cdot10^{21}\cdot\left(\frac{m_{\nu}}{1 \cdot \mathrm{eV}}\right)^{-1}$~eV \cite{Fargion1999}. The final nucleon secondaries may peak at $\simeq 10^{20}$~eV, as the maximal UHECR observed energies. Incidentally, let us note that this huge neutrino energy (transparent) transfer  in Z burst way may also help to lead TeVs energy showers in wide (Mpcs) neutrino halo, shower fed by Z boson fragments, as the pion secondaries photons at PeVs and soon later becoming by pair production  TeVs photons: this allow BL~Lac and AGN sources to overcome naturally the puzzling severe infrared TeV cut-off, somehow similar to the same GZK one \cite{Fargion(2004), Fargion_Oliva(2006)}. This  model  has been the very popular solution for the apparent over-GZK UHECR events since 1997-2004 period. However a few years later (1998-2007) the neutrino oscillation discoveries, (with cosmological constrains), were favoring a  little and lighter neutrino masses, maybe more comparable to the same tiny atmospheric neutrino mass splitting $\simeq 0.05$~eV, making more difficult the ideal (but not excluded) fine tuning neutrino mass at $\simeq 0.4$~eV,  a mass needed for better UHECR effective production.
 In addition such a too-light (atmospheric) neutrino masses have difficulties in forming dense gravitational $\nu$ clustering, making less  efficient the  $\nu+\bar{\nu}_{\mathrm{relic}}$ beam dump halos along were the ultra-relativistic Z boson would take place. The consequent UHE ZeV neutrino flux needed to require  very high fluency values, making the model less attractive. Moreover, since 2007, the AUGER early records and interpretations were  favoring at once the local group correlation with the first 26 UHECR events, making (apparently) obsolete the Z scattering \cite{Fargion1999} (also called Z burst \cite{Gelmini2004}), model in UHECR.
Nonetheless, as we shall discuss, the apparent rediscover of much UHECR events correlated with far GZK volume sources, the eventual Mrk~421 and 3C~454 correlation with small clustering of UHECR events, the $\simeq 1$~eV sterile neutrino resurgence, all of them offer new reason  for the resurgent  Z Shower model discussed below.
Finally, in the last decade (2007-2017) the early 2007 PAO \cite{PAO(2007)} claimed a correlation UHECR--SGP, that just faded away as soon as the more and more rich UHECR map was diluting and hiding any (early apparent) SGP statistical connection.
UHECR maps today show two main relic anisotropy, the so called Hot Spots (North and South) and a few narrow clustering in a noisy almost homogeneous sky. The most dramatic result, in our view, is the absence of the Virgo Cluster. These UHECR maps and anisotropy have found no astronomical meaning for most authors \cite{IceCube(2016)} or a  tentative correlations based on lightest UHECR courier  by us \cite{Fargion2016c}, correlation possible also with a few UHE highest energetic IceCube neutrino events.
 Let us now remind and critically comment the  early and the present most popular trend in the UHECR modeling and their motivations for a far edge UHECRs. The new popular fashion favor most distant UHECR sources at GZK edges (later on we shall review our different model based on lightest UHECR nuclei \cite{Fargion(2008), Fargion(2009)}.
 \begin{figure}[t]
\centering
\includegraphics[width=0.6\columnwidth]{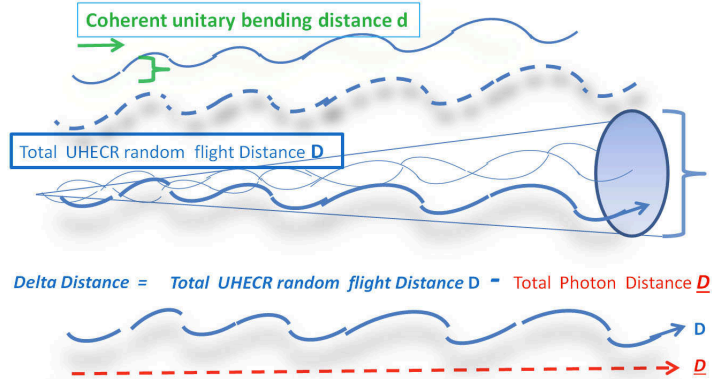}
\caption{A schematic  random walk of the far UHECR whose flight is bent and extended with respect to the straight flight for an AGN flaring photon; in this very simplified scheme please note the narrow unitary bending angle assumed as large as $1.75^\circ$ for a single coherent unitary bending of a distance $d_{\mathrm{Coh}}=1$~Mpc, $D_{\mathrm{Total}} =1$~Mpc, $B_{\mathrm{ExtraG}} =3$~nG; note also the final observed Hot Spot wide angle $\simeq17.5^\circ$ possibly due to the far flight: $D_{\mathrm{Total}}~=100$~Mpc. The difference  distances covered by the UHECR random walk and the direct photon flight define the time delay or the time lag discussed below.}\label{fig:1}
\end{figure}

\section{UHECR at GZK far edges?}

\begin{figure}[t]
\centering
\includegraphics[width = 0.70\columnwidth]{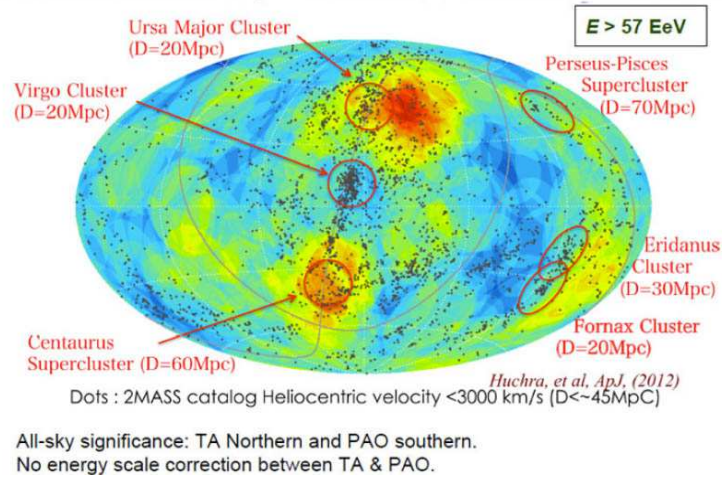}
\caption{The most recent UHECR Hot Spot, North by TA (telescope Array), South (by Auger) in celestial coordinate, shown in TA presentation slide. Note the central scandal of Virgo UHECR absence (not tag by red but by a poor dense blue colored area) at the UHECR map center, with sources candidate galaxies as  black dots point, for a maximal galaxy distribution within the best GZK 45~Mpc distances and volumes. There is a partial screening for Auger and also TA for the Virgo border line position, but its absence is anyway well in-written in the map.}\label{fig:2}
\end{figure}

Nowadays (2014-2017) both AUGER and TA detectors converged in two mild Hot Spot UHECR clustering (North and South, see Fig.\ref{fig:1}), while the most on fashion UHECR source candidates \cite{Kampert16, Deligny(2017)} for observed origination are 
the most distant AGN sources, at the edges of the allowed GZK cut-off radius ($\lesssim100$~Mpc), see Fig.~\ref{fig:3},~\ref{fig:4}, \emph{assuming a proton UHECR composition}.
Indeed most authors clearly argued the widest volumes may guarantee the observed tendency to a more and more homogeneous nearly isotropic UHECR sky. For them the two Hot Spots remain some unexplained event and maybe associated to some far galactic cluster. However we feel that while in balance of constrains some authors often underestimate the self consistence  of the whole UHECR transport model, including the Virgo absence for a proton currier and the clear UHECR composition shower signature toward light nuclei.
 The Virgo absence in UHECR sky is, even today, the most surprising result standing in both  PAO and TA data. Both detector arrays did not reveal any clustering on Virgo while this cluster  is the main largest and nearest (GZK volume) galaxy cluster source of infrared galaxy; its mass presence (nearly 2500 galaxies) is the  best observable sources within a GZK volume in infrared sky (IR): see Fig.~\ref{fig:2}.
 We remind that the GZK cut off volume for protons begins  about at $\simeq 50$~Mpc radius, as shown in Fig.~\ref{fig:2}; the GZK  is originated by the proton-photon scattering, making $\Delta$ resonances and photo-pion, somehow suppressing the UHECR flight distance. At a hundred Mpc there is already an initial exponential suppression for the highest energy UHECR, while at Virgo distance (20 Mpc) there is not any relevant GZK cut for protons: Virgo should be the best, nearest, brightest GZK sky \emph{(for proton UHECR)}. Therefore why Virgo cluster didn't reflected into AUGER and TA,  UHECR clusters, if the nucleons are the very main currier?
The UHECR Virgo absence in the expected UHECR map is not just a minor unexplained detail; it is instead (in our view) a major \emph{scandal}. It seems outrageous that on 2017 most authors just keep ignoring this huge embarrassing absence, see Fig.~\ref{fig:2}. UHECR GZK cut-off may be better allow to shine the Virgo cluster in place to any more far (a hundred Mpc)
clusters, that it is already partially suppressed by the distance dilution and by the initial exponential GZK opacity.
Nevertheless in the following section we shall follow the most recent proposal for UHECR correlation with
far hundred Mpc sources showing their probable inconsistence.

\subsection{Far UHECR bending path in random walk}

\begin{figure}[t]
\centering
\includegraphics[width = 0.70\columnwidth]{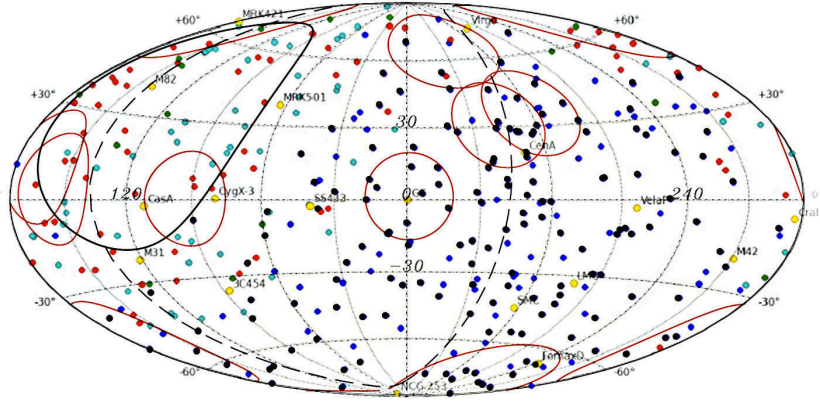}
\caption{The Galactic Hammer map with all published AUGER and TA as well as AGASA UHECR events
 with several proposed  red circle area $18^\circ$ wide, of far  (130~Mpc) AGN sources; the colored red circles are the TA events, while blue circles are the AUGER UHECR events;  the cyan ones show the older AGASA events \cite{Fargion(2015), Kampert16}}\label{fig:3}
\end{figure}

\begin{figure}[t]
\centering
\includegraphics[width = 0.70\columnwidth]{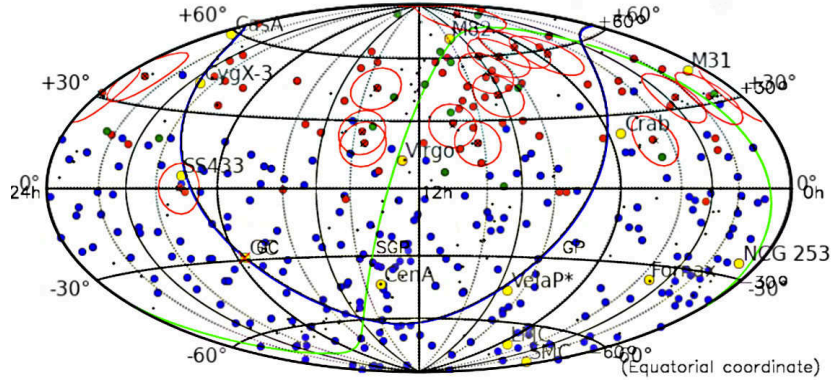}
\caption{The Celestial coordinate sky with all published AUGER and TA as well as AGASA UHECR events with several red circle area of far  ($z\simeq0.02$ or 80~Mpc) AGN sources.
 The sky plot (Aitoff projection, equatorial coordinates) of the TA UHECR events
with  $E_{\mathrm{UHECR}} > 62.2$~EeV (crosses) and the objects from the Swift BAT AGN catalog with redshift
$z<0.02$ (dots). Red circles around  the positions of UHECR events have a radius of $10^\circ$. Blue
and green lines show the Galactic and Super-galactic planes, respectively; the colored red circles are the TA events,
while blue circles are the AUGER UHECR events;  the cyan ones show the older UHECR AGASA events \cite{Fargion(2015), Deligny(2017)}}\label{fig:4}
\end{figure}

The observed PAO and TA  UHECR events are mostly spread and diffused homogeneously on the sky map. Nevertheless since a decade UHECR are showing a marked anisotropy at large scale angles ($\pm18^\circ$): the so  called twin hot spots. According to some authors (and us, among the earliest) they are originated by nearest AGN, Cen~A (Southern Hot Spot in PAO maps), and very possibly AGN M82 (Northern Hot Spot in TA data). In addition we also suggested that a very few and hard UHECR narrow clustering ($\pm5^\circ$) along the galactic plane might be originated by nearest (mostly galactic) sources; we foresaw correlated few UHE$\nu_\mu$ (above hundred TeV energy) and through-going $\mu$ (via their neutrino counterpart) in IceCube with such narrow clustering.
The UHECR wide Hot Spot might be also originated by more distant AGNs (a hundred Mpc) at the edge of  nucleon GZK cut off distances.
This opportunity is appealing to some authors because a hundred Mpc distance radius guarantees enough sources and homogeneity \cite{Kampert16, Deligny(2017)}; see Fig.\ref{fig:3}, \ref{fig:4}. However, the unique UHECR particles that may travel so far, as we mentioned are protons, while PAO models and simulation composition recently  favor a mix of lightest nuclei (He) together with light ones (Be, B, C and N), as we did suggested, independently,  since long time \cite{Fargion(2008), Fargion(2009)}, due to the problematic Virgo absence and the observed smearing angles.

 Let us now review the hypothesis of a far random walk ($\sim100$~Mpc) proton-smeared UHECR, assuming an UHECR with atomic number $Z$ (with respect to $Z_p=1$), energy $E$, total traveled distance $D_{Total} = D$ in extragalactic magnetic fields $B$ of unitary characteristic  coherence length $d_c$,  then the final random arrival angle $\alpha_{\mathrm{rm}}$ can be written as follows \cite{Fargion(2004)}:
\begin{equation}\label{angle}
\alpha_{\mathrm{rm}}=5.82^\circ\left(\frac{Z}{Z_p}\right)\left(\frac{E}{6\cdot10^{19}\,\mathrm{eV}}\right)^{-1}\left(\frac{D}{100\,\mathrm{Mpc}}\right)^{1/2}\left(\frac{d_c}{\mathrm{Mpc}}\right)^{1/2}\left(\frac{B}{\mathrm{nG}}\right).
\end{equation}

 Now, in order to be consistent with the observed UHECR smeared anisotropy Hot Spot angle around Cen~A of nearly $18^\circ$ we may better assume for 100~Mpc sources and averaged magnetic fields $B$ of about $3$~nG, so that eq.~(\ref{angle}) may be written as
\begin{equation}\label{anglep}
\alpha^p_{\mathrm{rm}}=17.5^\circ\left(\frac{Z}{Z_p}\right)\left(\frac{E}{6\cdot10^{19}\,\mathrm{eV}}\right)^{-1}\left(\frac{D}{100\,\mathrm{Mpc}}\right)^{1/2}\left(\frac{d_c}{\mathrm{Mpc}}\right)^{1/2}\left(\frac{B}{3\,\mathrm{nG}}\right).
\end{equation}

\section{Time delay of bent and smeared UHECR}

 The same angle may fit the Northern Hot Spot but the consequent flight time delay between a random walk UHECR and a direct photon flight would be
 \begin{equation}\label{time}
\Delta t\simeq\frac{D\alpha_{\mathrm{rm}^2}}{4c}\sim3.75\cdot10^6\left(\frac{E}{6\cdot10^{19}\,\mathrm{eV}}\right)^{-2}\left(\frac{D}{100\,\mathrm{Mpc}}\right)^{2}\left(\frac{d_c}{\mathrm{Mpc}}\right)\left(\frac{B}{3\,\mathrm{nG}}\right)^2\,\mathrm{yr}.
\end{equation}
This huge time lag, delay and dilution of the UHECR flight, see Fig.\ref{fig:1}, from AGN is  remarkable. Indeed, observed AGN held their $\gamma$ activity possibly for  centuries or thousand years, but possibly well below a million years \cite{Eilers(2017)}. Furthermore, their sudden flare blazing, as the hour long huge brightening from 3C~279, did eject an  energy flux comparable, in three days flashes, to a Gamma Ray Burst (GRB) (with typical power-energy of $\sim10^{49}$~erg/s); in general an active bright AGN should consume one or several solar mass in energy of the jet each year. It seem quite un-probable that such an energy supply may held for million years requiring such a huge an amount of sinking mass: an AGN in general it is not supposed to power along million years. Indeed the observed helical jet as in Mrk~501 \cite{1995ApJ...439...98C}, suggests two consequences: the thin AGN $\gamma$ jet is fueled by an accretion disk made by dense star cluster but the jet is often bent by a binary Black Hole or by an asymmetric  accretion disk, making thin jet in helical structures whose variabilities may be as short as minutes, hours or days, comparable or below  the same AGN-Black Hole, Schwarzschild scale-times, as in recent 3C 279 exceptional unexplained  flare\cite{Fargion_Oliva(2016)}.
Somehow the same allowed ages of the AGN lifetime is inscribed in the observed AGN radio or BL Lac x-gamma jet size: a few or tens kpc long suggesting a maximal age of a few hundred thousand years for AGN lifetime. In the same scenario one may include our gamma galactic Fermi bubble probably ejected along such a limited life time (below one million years).
The same spinning and precessing thin jet alignment with Earth may change so much with time that within many tens of thousand years, or a million years, it can be safely assumed that the $\gamma$ AGN flare is no longer apparently active, or at least pointing and blazing to us while being hit by their (much earlier ejected), now present UHECR.
Therefore, the same far  AGN sources proposed as the smoking gun for the UHECR \cite{Kampert16, Deligny(2017)}  ($\sim100$~Mpc, see Fig.~\ref{fig:3}, \ref{fig:4}), are probably well hidden and exhausted AGN because they were ending their accretion disk fuel, once they reached us millions years later. In brief there is little room in doing such a far AGN correlation with present gamma AGN and the very future late UHECR arrivals. Somehow in analogy to AGN precessing jet but at a smaller  mass scales let us remind the few-tens solar mass NS-BH tidal disruption, corresponding to a much shorter  time scales (milliseconds) feeding and precessing jet that we suggested in GRBs, \cite{Fargion(1999), Fargion_Oliva(2016b)}. Therefore we here reconsider and suggest two complete alternative approaches for the UHECR origination; being the second one the most conservative and realistic we would like to note that none of them excludes the other:


\begin{itemize}
\item Different view: UHECR by UHE ZeV neutrino; a far UHE ZeV neutrino extragalactic sources able to convert, via $\nu+\bar{\nu}\to Z\to\,$UHECR scattering on relic neutrino and via Z boson resonance their high energy into UHECR $p$, $\bar{p}$, $n$, $\bar{n}$ secondaries, overcoming the cosmic GZK cut off \cite{Fargion(1999)}: this guarantees a smooth isotropic noise and a possible Mrk~421 and 3C 454 correlation with UHECR. The Virgo absence is partially related to the overabundance of the far Z-shower cosmic sources.
\item Different view: The UHECR He, Li, Be, opacity; a nearby (a few Mpc distance) AGN and a few galactic UHECR sources made by light, mostly by lightest nuclei composition (He, Li, Be) and their isotopes.
\end{itemize}

  Let us conclude, briefly reminding, a very  relevant role of light or heavy UHECR nuclei whose boosted radioactivity in flight \cite{Fargion(2011)} it might reflect in the observed TeVs PeVs $\gamma$, $\beta$, $\alpha$ unexplained anisotropy in Milagro, ARGO and HAWC TeVs sky.

\section{Different view: UHECR by UHE ZeV neutrino}
\label{sec:1}

\begin{figure}[t]
\centering
\includegraphics[width = 0.70\columnwidth]{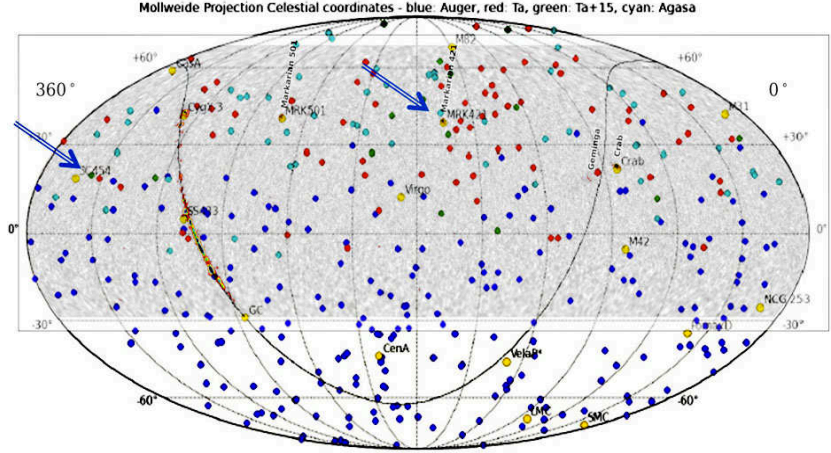}
\caption{UHECR events with their circled color, blue for AUGER, red for TA, and additional cyan for AGASA records \cite{Fargion(2015)};
note the tag shown by heavy red arrows pointing to brightest AGN and their nearby train of UHECR events. These correlations, if they will be confirmed, will imply a Z resonant connection where ZeV neutrino scatter relic ones \cite{Fargion(1999)}.}\label{fig:5}
\end{figure}

The UHECR map shows some narrow clustering also correlated in a train of events with very far active AGN, at a huge cosmic distances; indeed, as shown at the left side and the north sky in Fig. \ref{fig:5}, it might be that a group of UHECR are connected to the most powerful renowned Fermi source in $\gamma$, as AGN 3C 454, as well as Mrk 421.
However these AGNs are located much more above GZK cut off, up to  half a way across our universe, nearly two Giga-parsec. The most penetrating  UHECR nucleons are bounded, at best, as it is well known by cosmic Big Bang infrared noise, within fifty or a hundred Mpc \cite{GZK(1966)};  in this view one may suggest a meaningless coincident event; however a well known model \cite{Fargion(1999)} it is  able to overcome  the photon opacity at far distance well above GZK bound \cite{Fargion(1999), Fargion_Oliva(2006)} is  needed. It is, in fact, based on relic neutrino in dark hot halos offering the role of calorimeter  for UHE ZeV neutrinos whose scattering makes Z bosons.
In early articles we imagined a few eV relic neutrino mass able to fine tune the ZeV incoming energy leading to Z boson resonance.

However present limits on neutrinos are a fraction of eV mass or below; the very recent revival of interest for a sterile right-handed neutrino whose mass range at eV may offer a new favorite candidate particle for Z boson resonance.  Indeed once $\nu_{\mathrm{sterile}}$ became after cosmic expansion a non relativistic  such $\nu$ may behave as a left handed one as a usual active neutrino. Therefore neutrino pair annihilation may occur.
Then a far AGN above GZK distances may be an active source of UHECR if the UHE $\simeq 4 \cdot 10^{21}$ eV  neutrino scattering on relic ones may produce Z resonance (or a Z-Burst): its ultra relativistic decay in flight (within few Mpc dark neutrino halo)  may lead to nucleons and antinucleons tracks and they may overcome the very severe GZK cut off.
The result indeed, if it will be confirmed by additional meaningful crowding of UHECR toward far AGN (as Mrk 421, 3C 454)  it might be the first indirect discover of the UHE neutrino scattering on the relic neutrino whose mass may range in an allowable mass of a few eV up to $0.4$~eV  \cite{Fargion(2004), Fargion_Oliva(2006)}.
The presence of such a dark halo, if confirmed,  might be the most spectacular astrophysical road-map to a neutrino mass detection and its measure, as well as the best tool to reveal the largest neutrino clustering halos.
There is another model, more conventional, based on light UHECR nuclei within galactic and narrow universe that is not in contradiction with the present ZeV view nor mutually exclusive.

\section{Different view: The UHECR He, Li, Be, opacity}
\label{sec:2}

\begin{figure}[t]
\centering
\includegraphics[width = 0.8\columnwidth]{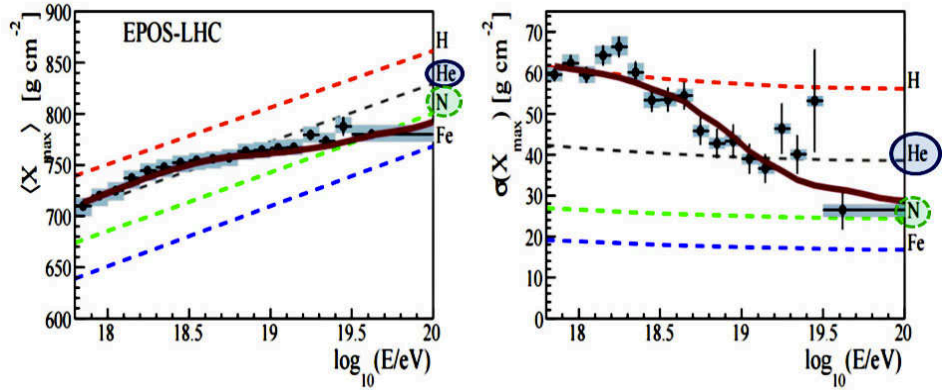}
\caption{The most recent UHECR composition signature with AUGER models \cite{PAO(2014)}.}\label{fig:6}
\end{figure}
\begin{figure}[t]
\centering
\includegraphics[width = 0.6\columnwidth]{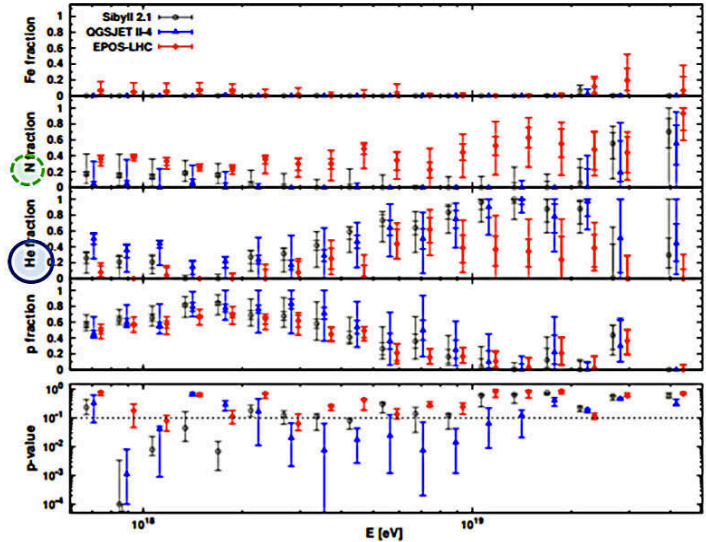}
\caption{The most recent simulation of UHECR air shower behaviors  whose  result confirm He (lightest) and  light nuclei like  (N) (or maybe Be, Li too). Note the almost missing proton and iron candidature in the maxima UHECR energies \cite{PAO(2014)} }\label{fig:7}
\end{figure}
\begin{figure}[t]
\centering
\includegraphics[width = 0.6\columnwidth]{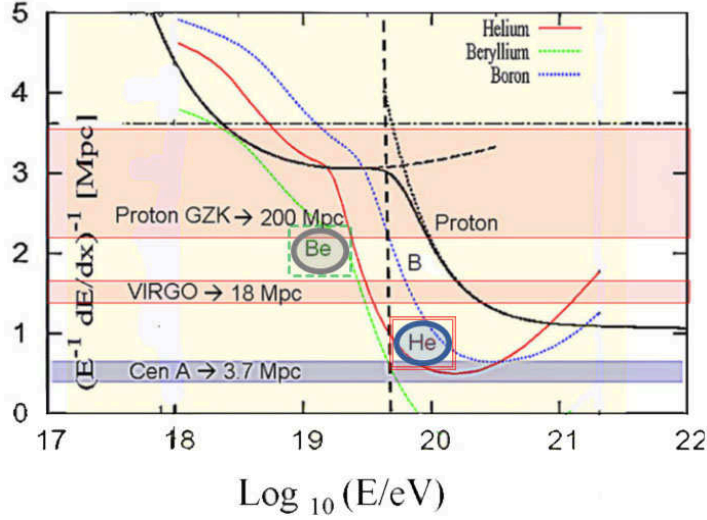}
\caption{The consequent allowed distance due to nuclear-photo-dissociation for lightest UHECR nuclei.}\label{fig:8}
\end{figure}

The absence of Virgo UHECR events calls for an answer.
Naturally, any light or lightest UHECR nuclei (He, D, T, Li, Be) being more fragile to cosmic $2.7$ K it is almost totally screened from Virgo by the consequent photo-nucleon dissociation in flight and the screen opacity.
Thus the lightest nuclei short flight may better face and explain the Virgo absence, see Fig.~\ref{fig:8}. It may also better coexist with PAO (and TA as well) light nuclei most recent composition signature, Fig. \ref{fig:6} and \ref{fig:7}, found in their fluorescence air-shower slant depth; indeed the most accurate estimate of the UHECR composition by AUGER is pointing to lightest (He) and light (N) composition \cite{PAO(2014)}, see Fig. \ref{fig:7}.
We remind that the N atom does not differ so much from the lighter nuclei as Be, whose role maybe also important in the missing Virgo signals.


The very popular (surviving even today) UHECR composition based on Fe was surely attractive, because the highest
nuclei may fit better with the highest energy for a given Lorenz factor. However  UHECR Fe are much more bent in our galaxy
because of the larger charge: 26 times more than a proton (whose bending is typically of about $3^\circ$),
leading to a total smearing or even to a galactic capture of such a heavy UHECR courier. Therefore, of course, there is no astronomy by Fe UHECR.
Iron, cannot merely explain any observed hot spot ($\pm20^\circ$) even in our galaxy. Finally, as we discussed, the UHECR protons, the second and main favorite UHECR courier, are arriving to us (by maximal far as $D=80$--$100$~Mpc) with a huge time delay or time lag, with respect to their present x-$\gamma$ AGN flaring activity. This time delay may disconnect the present $\gamma$ AGN flare activity with its (much later) UHECR arrival on Earth. On the contrary the very near bounded lightest nuclei UHECR sky (below a few Mpc) has a flight so short to be still consistent with most AGN survival lifetime. The flight time delay grows with the square of the distance making the 100 Mpc versus few Mpc three order of magnitude longer. Moreover UHECR protons courier, we underline,  would have to collimate and shine by huge nearest galaxy cluster sources (as IR galaxy clusters in GZK volumes) within a few degrees, clusters linked to Virgo, that as we often repeated, is totally absent, see Fig \ref{fig:2}.

\subsection{Light nuclei UHECR and their bending angle}

The present Hot Spot and their angular spread angle might be compared with light nuclei bending angle;
let us consider one of the light nuclei, Be, whose bending angle may extend up to:
\begin{equation}\label{anglegalBe}
\alpha^{gal}_{\mathrm{Be}}= 31^\circ\left(\frac{Z}{Z_{\mathrm{Be}}}\right)\left(\frac{E}{6\cdot10^{19}\,\mathrm{eV}}\right)^{-1}\left(\frac{D}{20\,\mathrm{kpc}}\right)^{1/2}\left(\frac{d_c}{\mathrm{kpc}}\right)^{1/2}\left(\frac{B}{3\,\mu\mathrm{G}}\right)
\end{equation}
This value is not far from the observed spread Hot spot one (even if originated just within our own galaxy).
The He, Li, Be  nuclei suffer of the photo-nucleon dissociation and cannot arrive from more than a few Mpc. Therefore their birth on Virgo is suppressed by such a fragile behavior of such nuclei. The most relevant one is the lightest one, the He.
Also D, Li, Be behave as well in opacity from Virgo. The Nitrogen, anyway,  is not totally screened from Virgo. However
 note that the Nitrogen (N) as an UHECR candidate (a light nuclei usually considered), does deflect nearly twice than the Be nuclei, leading to  a really wide spread angle, not much corresponding to the observed Hot Spot size.
\begin{equation}
\alpha^{gal}_{\mathrm{N}}=54.5^\circ\left(\frac{Z}{Z_{\mathrm{N}}}\right)\left(\frac{E}{6\cdot10^{19}\,\mathrm{eV}}\right)^{-1}\left(\frac{D}{20\,\mathrm{kpc}}\right)^{1/2}\left(\frac{d_c}{\mathrm{kpc}}\right)^{1/2}\left(\frac{B}{3\,\mu\mathrm{G}}\right)
\end{equation}

\subsection{Lightest nuclei UHECR and their time delay}

One of course may wonder if this is not the same case of the nearest AGN, as we did suggested since 2008 \cite{Fargion(2008)}, made by light and lightest nuclei \cite{Fargion(2009)}. In this case, indeed, the bending angle is scaling with the atomic number $Z$
($Z_{\mathrm{He}}=2$, $Z_{\mathrm{Li}}=3$, $Z_{\mathrm{Be}}=4$, $Z_{\mathrm{B}}=6$). Let us evaluate the two contributions for the extragalactic and in particular interest for the galactic bending angle:
\begin{equation}\label{angleex}
\alpha^{extra}_{\mathrm{He}}=1.64^\circ\left(\frac{Z}{Z_{\mathrm{He}}}\right)\left(\frac{E}{6\cdot10^{19}\,\mathrm{eV}}\right)^{-1}\left(\frac{D}{2\,\mathrm{Mpc}}\right)^{1/2}\left(\frac{d_c}{\mathrm{Mpc}}\right)^{1/2}\left(\frac{B}{\mathrm{nG}}\right)
\end{equation}
\begin{equation}\label{anglegal}
\alpha^{gal}_{\mathrm{He}}=15.5^\circ\left(\frac{Z}{Z_{\mathrm{He}}}\right)\left(\frac{E}{6\cdot10^{19}\,\mathrm{eV}}\right)^{-1}\left(\frac{D}{20\,\mathrm{kpc}}\right)^{1/2}\left(\frac{d_c}{\mathrm{kpc}}\right)^{1/2}\left(\frac{B}{3\,\mu\mathrm{G}}\right)
\end{equation}
which leads to a total of $\alpha^{extra}_{\mathrm{He}}+\alpha^{gal}_{\mathrm{He}}\simeq17.2^\circ$, in agreement with the Hot Spot spread angle. The time flight delay is mostly due to the extra galactic field; thus it can be evaluated as
 \begin{equation}\label{time2}
\Delta\tau\simeq6\cdot10^3\left(\frac{Z}{Z_{\mathrm{He}}}\right)^2\left(\frac{E}{6\cdot10^{19}\,\mathrm{eV}}\right)^{-2}\left(\frac{D}{2\,\mathrm{Mpc}}\right)^{2}\left(\frac{d_c}{\mathrm{Mpc}}\right)\left(\frac{B}{3\,\mathrm{nG}}\right)^2\,\mathrm{yr}
\end{equation}

\begin{figure}[t]
\centering
\includegraphics[width = 0.7\columnwidth]{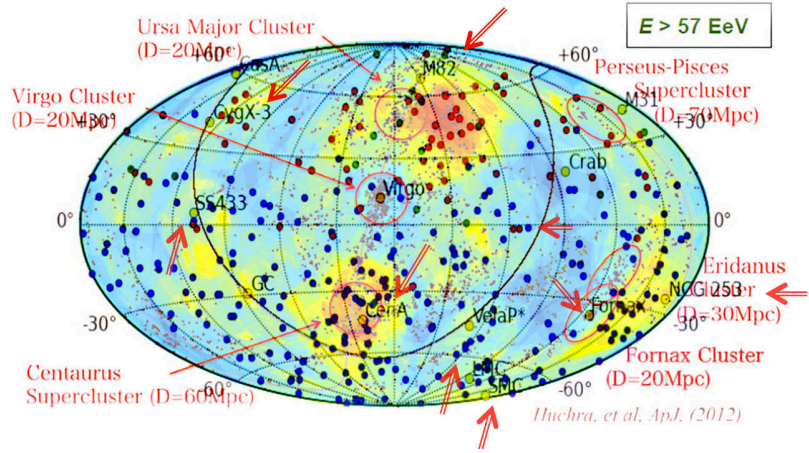}
\caption{As above the UHECR events with their circled color, blue for AUGER, red for TA, and additional AGASA cyan dots \cite{Fargion(2015)};
note the tag nearby sources shown by several short red arrows are claimed \cite{Fargion(2015)} to be the main origination sources of lightest nuclei UHECR}\label{fig:9}
\end{figure}

Have to be noted that both AGNs, Cen-A and M82, are not pointing or blazing to us right now. Their jet is so long (up to a few kpc) and extensive that it held activity possibly since several thousand years ago. On the contrary and as discussed above, a too far AGN ($\sim100$ Mpc) has a time lag of a few $10^6$ years (between UHECR and $\gamma$ arrival). The AGNs are expected to have a lifetime of $10^{5}$ yr, much less than a few million years of the time lag estimated above. Therefore no surprise that a UHECR may correlate with nearest (few Mpc) AGN as Cen~A, M82, NGC 253, while there are obvious doubts about possibility to correlate active AGN $\gamma$ source with much later signals coming hundred Mpc far away.

\section{EeVs anisotropy tracing LMC, Vela, Cen~A}
\begin{figure}[t]
\centering
\includegraphics[width = 0.7\columnwidth]{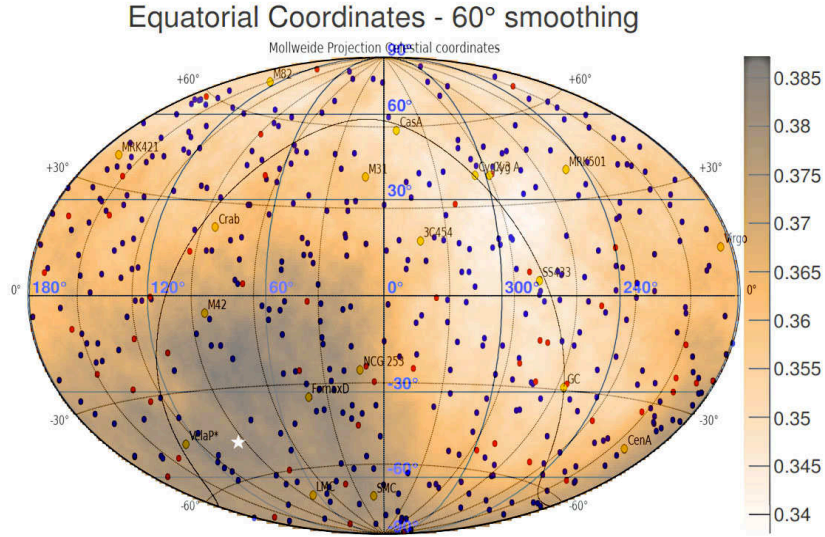}
\caption{Sky map in equatorial coordinates of the average UHECR flux reconstructed from data
recorded at the Pierre Auger Observatory and the Telescope Array above $10$ EeV smoothed
out at a $60^\circ$ angular scale, in km\textsuperscript{-2} yr\textsuperscript{-1}
sr\textsuperscript{-1} units. Note that present coordinate are rotated respect usual celestial coordinate. Under  transparently  we added the higher energy AUGER and TA and AGASA events UHECR as well their label names, tracing  several candidate source already considered \cite{Fargion(2015)}. Note in particular the Vela nearby train of UHECR events, LMC, SMC, Cen A, Fornax D, NGC 253, M42 sources }\label{fig:10}
\end{figure}

The recent review of the TA and the AUGER anisotropy of UHECR at $10$ EeV energies, as shown in Fig. \ref{fig:10}, smeared by $60^\circ$ reveals a surprising dipole anisotropy.
 No other deviation from isotropy has been observed at smaller angular scales. The recovered moment
can be visualized in Fig.\ref{fig:10}, where the average flux smoothed out at an angular scale of $60^\circ$
per solid angle unit is displayed using the Mollweide projection, in km\textsuperscript{-2} yr\textsuperscript{-1}
sr\textsuperscript{-1} units. Overlap is the main tag names of near galactic and nearby extragalactic sources.
This map is drawn in equatorial coordinates. The direction of the reconstructed dipole is
shown as the white star in Fig.\ref{fig:10}. It is obvious that at $10^{19}$ eV energy, the (eventual) proton UHECR courier,  will fly from a far extragalactic edges
being almost unconstrained by GZK cut-off, possibly making a more homogeneous and isotropic background sky signals. Such a dipole moment on the contrary is observed with an
amplitude $r_{\mathrm{UHECR}}$ = $6.5\,(\pm1.9\%)$, it has been captured with a chance probability of $5\cdot10^{-3}$ . To comparison the  eventual dipole due to the Sun (and our galaxy)
motion inside the Cosmic BBR has  a much smaller amplitude   $r_{\mathrm{BBR}}$ = $ \pm 0.2 \%$; moreover such anisotropy should be  pointing toward a quite different (Great attractor) sky. Therefore the most realistic scenario requires a nearby sources as the ones underwritten in the  Fig.\ref{fig:10}. Vela train of UHECR events, the clustering along LMC, SMC, Fornax D, NGC 253, and Cen A, (see in  Fig. \ref{fig:10})  all of them are among the possible root of these dipole tens EeV clustering  events: Indeed they are  contained in the main dipole shadow anisotropy (see in  Fig. \ref{fig:10}).
Therefore light UHECR nuclei or the lightest ones, may feed and justify the dipole anisotropy by their very local mass distributions.

\section{Conclusions}

We stand after a decade \cite{PAO(2007)} of pioneer AUGER   data   on our earlier interpretation;  UHECR are not correlated to any super-galactic plane, as last recent  data confirmed. Nor UHECR are a mixture of proton and iron courier (as many authors still believe), but, as we suggested  and we still confirm, UHECR are mostly light or lightest nuclei \cite{Fargion(2008), Fargion(2009)}, as the very updated  AUGER composition modeling (with correlated data) has probed \cite{Vitor2017}.
Indeed there is a very unique important signal driving to this conclusion: the UHECR Virgo cluster absence. This persistent lack of signals is the  most compelling argument for a light (lightest) UHECR composition. Indeed these light nuclei (D, He, Li, Be) cannot come from Virgo, because of their very short GZK cut off mode ($\sim$ Mpc) ruled by photo-nuclear dissociation and opacity. Their secondaries UHECR, at 20 EeV, fragments along the same Cen~A UHECR events, have been foreseen \cite{Fargion(2009b)} and they have been later observed in their foreseen shower signature \cite{PAO(2011), Fargion(2009b)}. The present UHECR mild clustering toward Cen-A and M82,  SOUTH and North Hot Spot (as well as a very possibly clustering by Cygnus X3, LMC, Vela, SS433, NGC 253 and Fornax D)  may be better understood as the bent tail of light nuclei UHECR events originated at nearby galactic or a few Mpc distances. The very recent tens EeV UHECR dipole anisotropy cannot find any natural explanation by far (hundred Mpc) mass distribution, but they may find a solution as shown in Fig \ref{fig:10} by local (Vela, LMC, SMC, NGC 253) sources. We remind that if UHECR arrival is clustering more and more toward Mrk 412, AGN 3C 454, at the  far edge Universe sources, the old Z boson resonance model,  made by ZeV neutrino scattering on relic ones, may also play a (maybe marginal) role.  There are additional tools in UHECR search able to disentangle UHECR nucleon from UHECR nuclei: the expected consequent EeV Tau neutrino, secondaries  of the (eventual) proton GZK cut off and pion decays in cosmic flight originated by longest oscillation distance flight;  these UHE EeV neutrino $\tau$ interacting inside the ground (skimming the Earth crust),  escaping above the Earth, they are followed by their long path $\tau$ flight and decay in air, leading, to remarkable, noise free and amplified tau air-shower \cite{Fargion(2000)} astronomy: the $\tau$ air-shower may be soon the key signature and the cleanest signal observable by AUGER \cite{PAO(2009)}, TA or  by near future large array Grand \cite{Grand(2017)} experiments probing UHECR GZK photo-pions traces. The same EeV $\tau$ neutrino air shower non detection (as the present AUGER limit shows), it is already a signal: it is  constraining more and more  the expectation of UHECR proton dominated UHECR courier, favoring therefore once again a light nuclei composition, whose secondaries arise at a lower neutrino energy (few tens PeV energy peak) and fluxes, within near future largest detector views for $\tau$ air-shower fluorescence lights (AUGER, TA) or better by Cherenkov flashes observable in  crown array telescope from space  (CHANT proposal \cite{CHANT(2017)} or top mountains  \cite{Fargion(2006a), ASHRA(2013)}) and by largest radio  traces \cite{Grand(2017)} at wide area on mountain sites. Therefore highest energy Tau neutrino astronomy maybe the near future purest and deeper trace toward the mysterious UHECR sources.
%
\section*{Acknowledgements}
The authors wish to thank the suggestions and the support of Prof. Marco Casolino, INFN, Rome 2,
whose discussions had clarified the writing of the article.


\begin{thebibliography}{99}%

\bibitem{GZK(1966)} K. Greisen, Phys. Rev. Lett. \textbf{16} 748 (1966). G.T. Zatsepin, V.A. Kuzmin, JETP Lett.  \textbf{4}  78 (1966). Zh. Eksp. Teor. Fiz.  \textbf{4} 114 (1966)
\bibitem{Elbert(1995)} J. W. Elbert, P. Sommers, Astrophys. J.  \textbf{441} 151 (1995)
\bibitem{1999ApJ...510L..91M} G.~A. Medina-Tanco  Astrophys. J., \textbf{510}, L91 (1999)
\bibitem{2000AGASA} N. Hayashida et al., Astron. J.  \textbf{120}  2190 (2000)
\bibitem{2007HiRes} P. Sokolsky, HiRes Collaboration, 30th International Cosmic Ray Conference \textbf{4}  451 (2007)
\bibitem{Kampert16} K.H. Kampert, The Pierre Auger Collaboration,  arXiv:1612.08188v1 (2016)
\bibitem{PAO16} A. Aab, et al., The Pierre Auger Collaboration, arXiv:1611.06812  (2016)
\bibitem{2016JCAP} IceCube, Pierre Auger \& Telescope Array Collaborations, JCAP01(2016), \textbf{2016}, 1, 037 (2016)
\bibitem{Fargion1999} D. Fargion, B. Mele, A. Salis,  Astrophys. J.  \textbf{517}  725 (1999)
\bibitem{Fargion(2004)} D. Fargion, A. Colaiuda, Nucl. Phys. B (Proc. Suppl.) \textbf{136}  256 (2004) \bibitem{Fargion_Oliva(2006)} D. Fargion,  P. Oliva, Nucl. Phys. Proc. Suppl. \textbf{165}  116  (2007)
\bibitem{Gelmini2004} G. Gelmini, G. Varieschi, T. Weiler,  Phys. Rev. D \textbf{70}, 113005, (2004)
\bibitem{PAO(2007)} The Pier Auger Collaboration, Science \textbf{318} 938 (2007)
\bibitem{IceCube(2016)} IceCube Collaboration, JCAP01, 037, (2016)
\bibitem{Fargion2016c} D. Fargion, P. Oliva, arXiv:1611.00079 (2016)
\bibitem{Fargion(2008)} D. Fargion, Phys. Scr. \textbf{78}  045901 (2008)
\bibitem{Fargion(2009)} D.Fargion, D. D'Armiento, P. Paggi, S. Patr\`i  Nucl. Phys. Proc. Suppl. \textbf{190} 162 (2009)
\bibitem{Deligny(2017)}  O. Deligny, K. Kawata, P. Tinyakov, arXiv:1702.07209;PoS ICRC 2015, 395 (2016)
\bibitem{Fargion(2015)} D. Fargion, G. Ucci, P. Oliva, P. G. De Sanctis Lucentini, EPJ Web of Conferences, \textbf{99} 08002 (2015)
 \bibitem{Eilers(2017)} A.C. Eilers et al.,  arXiv:1703.02539 (2017)
 \bibitem{1995ApJ...439...98C} J.~E. Conway, J.~M. Wrobel,  ApJ, \textbf{439} 98  (1995)
 \bibitem{Fargion_Oliva(2016)} D.Fargion, P. Oliva, PoS MULTIF15, 053 SISSA (2016);
 \bibitem{Fargion(1999)} D. Fargion, ; Astron. Astrophys. Suppl. Ser.138:507-508,(1999)
 \bibitem{Fargion_Oliva(2016b)} D. Fargion, P. Oliva, arXiv:1605.00177 (2016)
 \bibitem{Fargion(2011)} D. Fargion, , IOP, JPCS, 375, 052014; NIMA, A  \textbf{692}  174  (2012)
 \bibitem{PAO(2014)}A. Aab et al. (Pierre Auger Collaboration)Phys. Rev. D \textbf{90} 122006 (2014)
 \bibitem{Vitor2017} V. de Souza, The Pierre Auger Collaboration, arXiv:1701.06812 (2017).
 \bibitem{Fargion(2009b)} D. Fargion,  NIMA51778 PII: S0168-9002(10)01230-1, 2010
 \bibitem{PAO(2011)}Pier Auger Collaboration, 10.1016/j.astropartphys.2011.10.004, arXiv:1111.2472 (2011)
 \bibitem{Fargion(2000)} D. Fargion, Astr. Phys. J. \textbf{570}  909  (2002); D. Fargion et al. Astr. Phys. J. \textbf{613}   12851301 (2004)
 \bibitem{PAO(2009)} J. Abraham et al. (Pierre Auger Collaboration) Phys. Rev. D \textbf{79}, 102001 (2009)
 \bibitem{Grand(2017)} O. Martineau-Huynh, Giant Radio Array for Neutrino Detection; arXiv:1702.01395 (2017)
 \bibitem{PAO(2008)} J. Bluemer, Pier Auger Collaboration, 10.1143/JPSJS.78SA.114 (2008)
 \bibitem{2016arXiv160203497F}  D. Fargion,,\& Oliva, P., Nucl. and Part. Phys. Proc., \textbf{279--281} 198 (2016)
 \bibitem{1999AAS..138..507F} D. Fargion, A\&A, \textbf{138} 507 (1999);  D. Fargion, M. Grossi, Nuovo Cim.C\textbf{28}:809-812,(2005)
  \bibitem{Fargion(2006a)} D. Fargion et al. Adv. Space Res. \textbf{37} (2006);  D.Fargion, et al. NIM A \textbf{588} 146150 (2008).
  \bibitem{ASHRA(2013)} Y. Asaoka, M. Sasaki,  ASHRA Collaboration, Astr Phys.  \textbf{41} 716  (2013)
 \bibitem{CHANT(2017)} A. Neronov et al., Phys. Rev. D. \textbf{95} 023004 (2017)



\end{thebibliography}
\end{document}